\documentstyle[aps,prl,multicol,epsfig,amssymb,amstex,pifont]{revtex}

\begin{document}

\title{Surface Acoustic Wave induced Transport in a Double Quantum Dot}

\author{W.J.M. Naber$^{1,2,3}$, T. Fujisawa$^{2,4}$, H.W. Liu$^{2,5}$ and W.G. van der Wiel$^{1,6}$}
\address{$^1$SRO NanoElectronics, MESA$^+$ Institute for NanoTechnology, University of Twente,\\
 PO Box 217, 7500 AE Enschede, The Netherlands}
\address{$^2$NTT Basic Research Laboratories, NTT Corporation, 3-1 Morinosato-Wakamiya, Atsugi, Kanagawa 243-0198, Japan}
\address{$^3$Kavli Institute of NanoScience, Delft University of Technology, PO Box 5046, 2600 GA Delft, The Netherlands}
\address{$^4$Tokyo Institute of Technology, 2-12-1 Okayama, Meguro-ku, Tokyo 152-8551, Japan}
\address{$^5$SORST-JST, 4-1-8 Honmachi, Kawaguchi, Saitama 331-0012, Japan; National Laboratory of Superhard Materials, Institute of Atomic and Molecular Physics, Jilin University, Changchun 130012, China}
\address{$^6$PRESTO-JST, 7-3-1, University of Tokyo, Hongo, Bunkyo-ku, Tokyo 113-8656, Japan}

\maketitle

\begin{abstract}
We report on non-adiabatic transport through a double quantum dot
under irradiation of surface acoustic waves generated on-chip. At
low excitation powers, absorption and emission of single and
multiple phonons is observed. At higher power, sequential phonon
assisted tunneling processes excite the double dot in a highly
non-equilibrium state. The present system is attractive for studying
electron-phonon interaction with piezoelectric coupling.
\vspace{-0.75cm}
\end{abstract}

\pacs{73.23.Hk,63.20.Kr,77.65.Dq}

\begin{figure}[htbp]
 \vspace{-0.5cm}
   \begin{center}
   \centerline{\epsfig{file=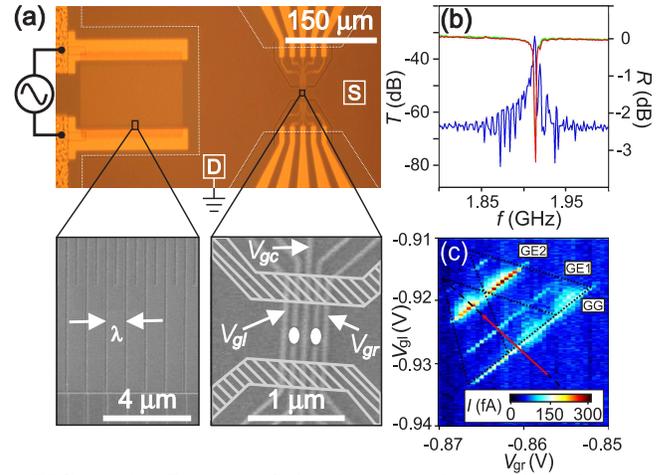, width=8.5cm, clip=true}}
     \caption{(a) Picture of the device with interdigital transducer
     (IDT, left) and double quantum dot (DQD, right). The source (S)
     and drain (D) reservoirs are indicated. The IDT-DQD distance is
     227.5 $\mu$m. In the scanning electron micrograph (SEM) of the
     IDT, the electrodes, separated by $\lambda = 1.4~\mu$m, are visible. In the
     hatched regions of the DQD SEM the 2DEG is depleted by shallow
     dry etching. The position of the dots is indicated by white dots.
(b) Transmission $T$ (blue curve) and reflection $R$ (red and green
curves) at room temperature of two IDTs similar to the one used in
the experiments, separated by a distance of 455 $\mu$m. A clear peak
in $T$ and a dip in $R$ are visible at 1.92 GHz. (c) Color scale
plot of the DQD current as a function of gate voltages $V_{gl}$ and
$V_{gr}$ at source drain voltage $V_{SD}=500$ $\mu$V without SAW
generation. The conductance triangles are accentuated by dotted
lines. Resonant tunneling lines are clearly visible. The dual gate
sweep direction for the SAW experiments is indicated by the red
arrow.}
     \label{device}
   \end{center}
   \end{figure}

\begin{figure}[htbp]
 \vspace{-0.5cm}
   \begin{center}
   \centerline{\epsfig{file=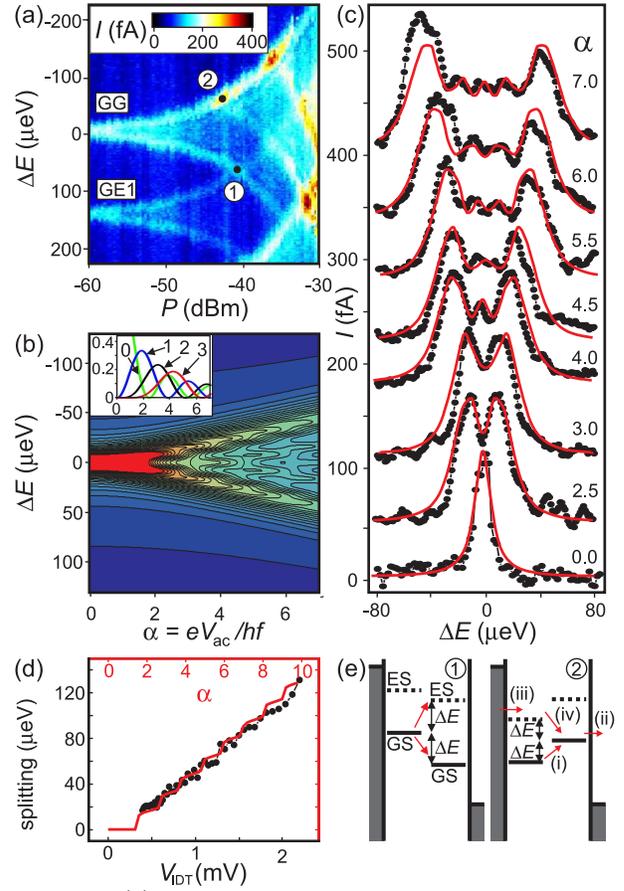, width=8.5cm, clip=true}}
     \caption{(a) Color scale plot of the DQD current versus $\Delta E$
     and microwave power $P$, at $f_{\rm{SAW}}=1.9446$ GHz, for the same transitions as in Fig$.$ 2.
     (b) Experimental (black dots) and calculated (red curves) current
spectra for different microwave powers, extracted from (a) and (c),
respectively. The experimental microwave power incident on the IDT
is converted to normalized potential amplitude $\alpha$ using (d).
The current height of the calculated spectra is fitted to the
experimental data. (c) Calculated DQD current versus $\Delta E$ and
$\alpha$ in the non-adiabatic limit, as explained in the text.
Inset: squared Bessel functions $J_{n}^{2}(\alpha)$ for $n$ = 0,1,2
and 3. (d) Splitting of the current peaks as function of the
amplitude of the microwave voltage $V_{\rm{IDT}}$ applied to the IDT
for the experimental data (black data points and axes), and current
peak splitting derived from the calculated spectra in (c) as
function of $\alpha$ (red curve and axes). By matching the
experimental and calculated curves, the conversion between $P$ and
$\alpha$ is found. (e) Schematic energy level diagrams for the
positions 1 and 2 indicated in (a). The transitions (i)-(iv) are
discussed in the text.}
     \label{powerdep}
   \end{center}
\vspace{-0.5cm}
    \end{figure}

\begin{multicols}{2}
Electron-phonon coupling often leads to dissipation and decoherence
problems in nanoelectronic devices. The decoherence in a tunable
two-level quantum system (qubit), such as a double quantum dot (DQD)
\cite{Wilfred03}, is of particular interest in the recent light of
quantum computation and information \cite{Leggett87}. It was found
that piezoelectric coupling to acoustic phonons is the dominant
mechanism for inelastic transition between two charge states in a
DQD \cite{ToshiSpont}, as confirmed by theory \cite{Brandes}. In
analogy to quantum states in natural atoms -- which dominantly
couple to, and are successfully controlled by photons -- the
electronic states in solid state systems may be controlled by
phonons, taking advantage of the strong electron-phonon coupling.\\
\indent Due to the piezoelectric coupling in GaAs, surface acoustic
waves (SAWs) can be generated by applying a microwave signal to an
interdigital transducer (IDT) \cite{Oliner}. The accompanying
propagating and oscillating potential has been used in several
experiments to transport photo-generated electrons and holes in
so-called `dynamical quantum dots' \cite{dynamicaldots}. In those
experiments, however, the SAWs give rise to an adiabatic change of
the electronic states, where the carriers remain in an eigenstate of
the temporal potential.\\
\indent In this Letter, we present {\it non-adiabatic} transitions
in a lithographically defined DQD under irradiation of coherent
SAWs. We observe resonant phonon assisted tunneling, where transport
is well described by considering absorption and emission of one or
multiple phonons during the tunneling process \cite{Stoof}. The
present results unambiguously indicate a finite contribution of SAWs
to the bosonic environment of a quantum two-level system formed by a
DQD. Moreover, these transport measurements allow us to determine
extremely small amplitudes of the local piezoelectric potential.\\
\indent Figure \ref{device}(a) is a picture of our device showing
the Ti/Au gate patterns of the interdigital transducer used for
generating SAWs on the left, and the DQD on the
\noindent right on top of a GaAs/AlGaAs heterostructure with a 2D
electron gas (2DEG) 100 nm below the surface. The periodicity of the
IDT is 1.4 $\mu$m, setting the SAW wavelength $\lambda_{SAW}$, and
corresponding to a SAW frequency of about 2 GHz in GaAs [see lower
left SEM micrograph in Fig$.$ \ref{device}(a)]. The IDT design is
characterized at room temperature using a different GaAs/AlGaAs
heterostructure with two identical IDTs facing each other, allowing
for a two-channel microwave measurement. The transmission and
reflection spectra in Fig$.$ \ref{device}(b) show a clear resonance
at 1.92 GHz, as expected from the IDT design. The reflection dip is
more than 3 dB, indicating that more than half of the incident power
is absorbed in the IDT. The transmission reaches a maximum of -30 dB
at resonance, implying additional loss in the device. Possible
mechanisms for power loss are impedance mismatch, electromechanical
conversion loss and Bragg reflection within the IDT. We found that
the reflection and transmission spectra do not change when a DQD
device is fabricated in the middle between the IDTs. By assuming
identical characteristics for both IDTs, acoustic power at the site
of the DQD
is 15 dB less than the incident microwave power, $P$.\\
\indent The DQD is formed in an etched channel of 600 nm width [see
hatched dry etching regions in the lower right SEM micrograph in
Fig$.$ \ref{device}(a)] with appropriate voltages to the indicated
gate electrodes, which have a 220 nm spacing \cite{Wilfred03}.\\
%
%
\indent All measurements described below, are performed in a
dilution refrigerator with a base temperature of 50 mK. We have
obtained similar results in two different samples, measured in
different cryostats. The data shown here, are taken from one sample.
Each dot contains $\sim$10 electrons, has a charging energy of
$\sim$2 meV and a discrete energy level spacing of $\sim$150
$\mu$eV. The inter-dot electrostatic coupling is $\sim$200 $\mu$eV,
and the tunneling coupling is weak ($\ll 10$ $\mu$eV) so that
delocalization of states can be neglected. This weak coupling regime
is suitable for studying electron-phonon interaction \cite{ToshiSpont}.\\
\indent Figure \ref{device}(c) shows the single-electron tunneling
current through the DQD versus gate voltages $V_{\rm{gl}}$ and
$V_{\rm{gr}}$ with a large bias voltage of 500 $\mu$V with no
microwave power ($P=0$) applied to the IDT. The lower and upper
(partly overlapping) triangular conduction regions correspond to
electron-like and hole-like transport through the DQD, respectively
\cite{Wilfred03}. Resonant tunneling through the ground states (GSs)
of the two dots corresponds to the current peak at the base of the
triangles (labeled GG), while other resonant tunneling between the
left GS and the first and second excited states of the right dot are
also observed (labeled GE1 and GE2). In the following measurements
we simultaneously sweep $V_{gl}$ and $V_{gr}$ along the red arrow in
Fig$.$ \ref{device}(c), so that the energy difference $\Delta E =
E_1 - E_2$ between the GS energies of the left dot ($E_1$) and the
right dot ($E_2$) is varied. We observe a symmetric current profile
around $\Delta E = 0$ representing elastic current through the DQD,
while inelastic current at $\Delta E > 0$ associated with
spontaneous emission of phonons is very small in the present
experiment.
\begin{figure}[htbp]
 \vspace{-0.5cm}
   \begin{center}
   \centerline{\epsfig{file=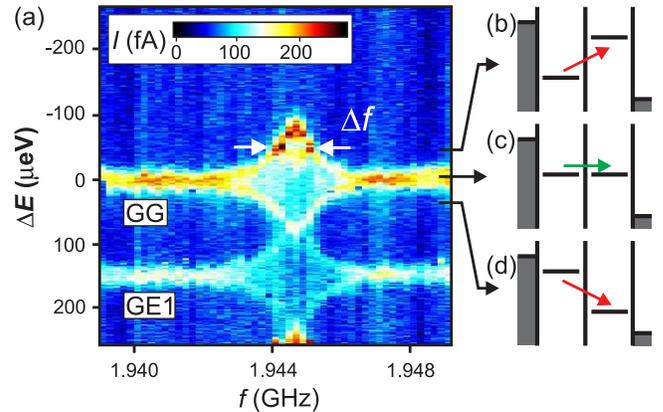, width=8.5cm, clip=true}}
     \caption{(a) Color scale plot of the DQD current versus ground state
     level spacing $\Delta E$ and microwave frequency $f$ applied to
     the IDT (-40 dBm microwave power). $V_{\rm{gl}}$ and $V_{\rm{gr}}$ are swept
     along the red arrow indicated in Fig$.$ 1(c). The current at $\Delta E=0$ and 150
     $\mu$eV corresponds to resonant tunneling through the ground states
     (GG), and through the left ground state and an excited state in the
     right dot (GE1), respectively. A clear resonance is observed at
     1.9446 GHz ($\Delta f=1.4$ MHz), corresponding to the IDT resonance
     frequency. The inelastic current is due to absorption and emission
     of SAW phonons, as schematically depicted in the energy diagrams (b)
     and (d), respectively. The energy diagram for elastic resonant tunneling is shown in (c).}
     \label{resonance}
   \end{center}
    \end{figure}
When microwaves are applied to the IDT, we observe significant
broadening and splitting of the resonant tunneling peaks only at the
IDT resonant frequency, $f_{\rm SAW} = 1.9446$ GHz, as seen in the
frequency dependence of the current spectrum in Fig$.$
\ref{resonance}(a). The resonance frequency corresponds very well to
that of the GaAs reference sample (1.92 GHz) of Fig$.$
\ref{device}(b), where the slight deviation is ascribed to the
different heterostructure and the lower temperature in the actual
device. This good correspondence rules out photon assisted tunneling
\cite{PAT}. There is no reason why there should be an
electromagnetic resonance coinciding with the IDT resonance
frequency. We also exclude resonant heating, since the energy levels
are well separated from the Fermi levels of the leads. The harmonic
oscillation of the energy levels as described below, cannot be
explained in terms of heating either. Note that no broadening is
observed at off-resonant frequencies, also indicating that heating
and spurious
electromagnetic coupling are negligible.\\
\indent We now look in more detail at the mechanism of the
SAW-induced current in Fig$.$ \ref{resonance}(a). The traveling SAW
causes a time-dependent potential $V_{\rm ac} \cos(2 \pi f_{\rm
SAW}t)$ between the two quantum dots, due to the piezoelectric and
deformation coupling. For GaAs at this frequency, the piezoelectric
effect is dominant and the deformation coupling can be neglected
\cite{Paulo}. As the lithographical dot-dot distance is $d$ = 220 nm
and the SAW wavelength is $\lambda_{SAW}$= 1.4 $\mu$m, $V_{\rm ac}$
is a fraction of the amplitude of the piezoelectric potential
$V_{\rm pe}$, $V_{\rm ac} = \eta V_{\rm pe}$, where $\eta = sin(\pi
d/\lambda_{SAW}) \approx$ 0.47. The time-dependent level spacing
$\Delta \widetilde{E}(t)$ is therefore $\Delta E + \eta V_{\rm
pe}\cos(2 \pi f_{\rm SAW}t)$. The peak splitting at resonance
frequency in Fig$.$ \ref{resonance}(a) can then be explained by a
propagating SAW in the adiabatic limit as follows. Energy diagrams
for positive $\Delta E$, $\Delta E =0$, and negative $\Delta E$ are
shown in Figs$.$ \ref{resonance}(b), (c), and (d), respectively.
Elastic current now appears at the time-dependent resonant condition
$\Delta \widetilde{E}(t)=0$. By assuming that the current is simply
proportional to the time spent at resonant condition, current peaks
appear at $\Delta E \pm \eta V_{\rm pe}$ (i$.$e$.$ at the classical
turning points). The peak splitting is hence related to the
piezoelectric potential amplitude.\\
\indent The microwave power dependence of the current spectra is
presented in Fig$.$ \ref{powerdep}(a). The peak splitting clearly
increases with microwave power $P$. In Fig$.$ \ref{powerdep}(d) the
splitting is plotted (black dots) as function of the amplitude of
the microwave voltage applied to the IDT, $V_{\rm{IDT}}$, confirming the linear dependence \cite{VIDT}.\\
\indent Since the tunneling rate (about 1 MHz for 100 fA current in
our weakly-coupled DQD) is much smaller than $f_{\rm{SAW}}$, an
electronic state in one dot acquires a phase, which is given by the
integration of the oscillating potential, relative to another state
in the other dot \cite{TienGordon63}. This non-adiabatic effect
appears for example as photon assisted tunneling, as evidenced in
various devices under microwave or far-infrared irradiation
\cite{PAT}. In our case, the oscillating potential is obviously
induced by {\it phonons}. One can say that the DQD is exposed to
surface acoustic phonons with energy $hf_{\rm SAW}$ = 8 $\mu$eV. The
energy-dependent tunnel rate $\widetilde{\Gamma}(E)$ from the left
dot to the right dot in the presence of the phonon field is given by
the same theory \cite{Wilfred03,TienGordon63}
\begin{equation}
\widetilde{\Gamma}(\Delta
E)=\sum_{n=-\infty}^{\infty}J_{n}^{2}(\alpha)\Gamma(\Delta
E+nhf_{\rm SAW}), \label{eq:tunnelrates}
\end{equation}
where $n = 0,\pm 1,\pm 2,...$ is the number of phonons involved in
the emission (positive $n$) and absorption (negative $n$),
$\Gamma(E)$ is the tunnel rate without phonons and
$J_{n}^{2}(\alpha)$ is the squared $n$-th order Bessel function of
the first kind evaluated at normalized amplitude $\alpha=eV_{\rm
ac}/hf_{\rm SAW}$ [see inset to Fig$.$ \ref{powerdep}(c)]. The
modulated DQD current $\widetilde{I}$ then becomes \cite{Stoof}
\begin{equation}
\widetilde{I}=e|t_{12}|^{2}\Gamma_{R}\sum_{n=-\infty}^{\infty}\frac{J_{n}^{2}(\alpha)}{\Gamma_{R}^2/4+(n2\pi
f_{\rm SAW}-\Delta E/h)^{2}}, \label{eq:current}
\end{equation}
where $|t_{12}|$ is the modulus of the tunnel coupling between the
two dots, and $\Gamma_{R}$ the tunnel rate from the right dot to the
right lead. Inelastic current is allowed whenever the level spacing
equals an integer number times the phonon energy, i$.$e$.$ $\Delta E
= nhf_{\rm SAW}$. The current thus consists of a number of satellite
peaks, separated by the phonon
\noindent energy $hf_{\rm SAW}$. The Bessel function describes the
probability that an electron absorbs ($n>0$) or emits ($n<0$) $n$
phonons. It should be noted that Eq$.$ (\ref{eq:current})
approaches the adiabatic limit for $\alpha \gg 1$.\\
\indent Our DQD device has a resonant current line width of 14
$\mu$eV, even at zero microwave power, which is not sufficient to
resolve phonon sideband with spacing $hf_{\rm SAW}$ = 8 $\mu$eV
\cite{Pioro}. However, we do observe clear evidence of non-adiabatic
effects in the current spectra, as described below. Figure
\ref{powerdep}(b) shows the DQD current as function of $\Delta E$
for different $P$. The lowest curve is measured at zero power and
represents the elastic current. The fit of the elastic current,
$I_{\rm{el}}$, (red curve) is a Lorentzian with a FWHM of 14
$\mu$eV. The expected current at finite microwave power, $I_{\rm
SAW}(\Delta E)$, is now derived from the zero-power curve as $I_{\rm
SAW}(\Delta E)= \sum_{n=-\infty}^{\infty} J_{n}^{2}(\alpha)I_{\rm
el}(\Delta E+nhf_{\rm SAW})$, and is plotted versus $\alpha$ in
Fig$.$ \ref{powerdep}(c). For $\alpha \gtrsim 2$ the resonant
current splits in two peaks whose positions approach $\Delta E = \pm
\alpha hf_{\rm{SAW}}$, corresponding to the adiabatic limit. The
splitting between the calculated current peaks versus $\alpha$ is
plotted (red solid curve) together with the experimental splitting
versus $V_{\rm{IDT}}$ in Fig$.$ \ref{powerdep}(d). Very good
agreement between the experimental data and the non-adiabatic
calculation is found when we relate $V_{IDT}$ to the normalized ac
potential $\alpha$ according to $\alpha=0.09 \eta eV_{IDT}/hf$. The
first factor, corresponding to the loss in the IDT, is in good
agreement with the loss estimated from Fig$.$ 1b.
\indent The non-adiabatic calculation in Fig$.$ \ref{powerdep}(c)
shows clear additional structure in between the split peaks. This
structure originates from the phonon satellite peaks that should be
individually resolvable at $\Delta E = nhf_{\rm{SAW}}$ if the peak
width is smaller than the phonon energy. In our case, however, the
peak width exceeds $hf_{\rm{SAW}}$ (but is less than
$2hf_{\rm{SAW}}$). We actually find good agreement between the
calculated current spectra and the experimental data (including the
inter-peak fine structure) at finite microwave power as shown in
Fig$.$ \ref{powerdep}(b), where we have applied the
$\alpha$-$P[\rm{dBm}]$ conversion derived in Fig$.$
\ref{powerdep}(d). Our data thus reveal clear quantum behavior, even
when we cannot resolve individual phonon satellites.\\
\indent Quantum behavior is also observed in multiple excitation
processes between excited states at higher power. As indicated by
\ding{172} in Fig$.$ \ref{powerdep}(a), one of the split peaks of
the GG resonance and one of the GE1 resonance touch around -40 dBm,
where the GS-GS level spacing and the spacing between the GS in the
left dot and the first ES in the right dot both equal $V_{\rm ac}$,
as shown in the left diagram of Fig$.$ \ref{powerdep}(e). At this
condition, two phonon assisted tunneling processes (red arrows) are
allowed from the GS of the left dot. There is another peak emerging
for $P > -42$ dBm indicated by \ding{173} in Fig$.$
\ref{powerdep}(a). This peak is associated with phonon assisted
tunneling from an ES in the left dot to the GS of the right dot
[right diagram of Fig$.$ \ref{powerdep}(e)]. This tunneling process
is only possible if an electron first tunnels from the GS of the
left dot to the GS of the right dot (i) and escapes to the right
lead (ii), another electron tunnels into the ES of the left dot
(iii) followed by tunneling to the GS of the right dot under phonon
emission (iv). This explanation is consistent with the absence of
the peak at lower power, where the GS in the left dot is permanently
occupied. At higher power, more resonant peaks are resolved, which
may be useful in analyzing the energy spectrum of our DQD.

Finally, we comment on the measurement sensitivity to the
piezoelectric potential in our experiment. As discussed above, the
current spectra reflect the amplitude of the local piezoelectric
potential. The lowest power at which we can resolve peak splitting
is -58 dBm, corresponding to $V_{\rm pe}$ = 24 $\mu$V, which is
several orders of magnitude smaller than the power used to induce
dynamical quantum dots \cite{dynamicaldots} and to induce lattice
displacements measurable by optical interferometry
\cite{MMdeLima03}. The minimum detection power can be improved
further by adjusting the DQD parameters. When the elastic current
peak width is made smaller than the phonon energy, the piezoelectric
potential can be derived from the amplitude of the phonon assisted
tunneling current via the Bessel function dependence even for
$\alpha \ll 1$. This may enable the measurement of lattice
distortion due to vacuum fluctuations.\\
%
\indent In conclusion, we have observed inelastic tunneling in a DQD
two-level system coupled to a monochromatic SAW source. The
transport through the DQD is well described by non-adiabatic
Tien-Gordon theory for resonant tunneling between two discrete
states with a time-dependent potential. We find that the DQD can be
employed as a very sensitive SAW detector and is promising for
studying electron-phonon interaction.

\indent We thank S. Tarucha, P.V. Santos, R. Aguado, L.P.
Kouwenhoven and Y. Hirayama for fruitful discussions and help. We
acknowledge financial support from DARPA grant number
DAAD19-01-1-0659 of the QuIST program, and SCOPE from the Ministry
of Internal Affairs and Communications of Japan.

 \end{multicols}
\end{document}